\documentclass{amsart}
\usepackage{amsmath}
\usepackage{amssymb}
\usepackage{latexsym}
\usepackage{amsfonts}
\usepackage[dvips]{graphicx}
\usepackage{color}
\usepackage{graphicx,epstopdf}

\newtheorem{conj}{Conjecture}

\begin{document}
\title
{zeroes of Wronskians of Hermite polynomials and Young diagrams}%
%
\author{G. Felder}
\address{Department of Mathematics, ETH Zurich, 8092 Zurich, Switzerland}
\email{felder@math.ethz.ch}
\author{A.D. Hemery}
\address{Department of
Mathematical Sciences,
          Loughborough University,
Loughborough,
          Leicestershire, LE11 3TU, UK}
          \email{A.D.Hemery@lboro.ac.uk}
          
         \author{A.P. Veselov}
\address{Department of
Mathematical Sciences,
          Loughborough University,
Loughborough,
          Leicestershire, LE11 3TU, UK
          and
          Moscow State University, Moscow 119899, Russia}

\email{A.P.Veselov@lboro.ac.uk}

\maketitle

\centerline{\it To Boris Dubrovin on his
$60^{\text{th}}$ birthday}

\begin{abstract} For a certain class of partitions, a simple qualitative relation is observed between the shape of the Young diagram and the pattern of zeroes of the Wronskian of the corresponding Hermite polynomials. In the case of two-term Wronskian $W(H_n, H_{n+k})$ we give an explicit formula for the asymptotic shape of the zero set as $n \rightarrow \infty$.  Some empirical asymptotic formulas are given for the zero sets of three and four-term Wronskians.
\end{abstract}



\medskip

\normalsize




\section{Introduction}

Consider a Schr\"odinger operator 
$$ L = -\frac{d^2}{dz^2} + u(z)$$
with a rational potential $u(z),$ not necessarily real.
Such an operator is called {\it monodromy-free} 
if all the solutions of the corresponding Schr\"odinger equation
$ L\psi = \lambda \psi $
are meromorphic in the whole complex plane for all $\lambda.$ 

The first classification result here is due to Duistermaat and Gr\"unbaum \cite{DG}, who described all monodromy-free operators with rational potentials decaying at infinity.  The pole configurations of the corresponding potentials were studied earlier by Airault, McKean and Moser \cite{AMM} in relation to rational solutions of the KdV equation.

Oblomkov \cite{O} generalised Duistermaat-Gr\"unbaum's result to the quadratic growth case. He showed that all the rational monodromy-free operators with rational potentials growing as $z^2$ at infinity are the results of Darboux transformations applied to the harmonic oscillator. The corresponding potentials have the form
$$u(z)=- 2 \frac{d^2}{dz^2} \log W (H_{k_1},\dots, H_{k_n}) + z^2 +c,$$
where $H_k(z)$ is the $k$-th Hermite polynomial, $k_1>k_2>\dots >k_n$ is a
sequence of different positive integers and $W (f_1,\dots,f_n)$ is the Wronskian of functions $f_1, \dots, f_n.$ 

We are interested in the geometry of the pole configurations of the corresponding potentials ({\it locus} in the terminology of Airault, McKean and Moser), which are the same as the zero sets of the corresponding Wronskians.
This locus has an interesting relationship with the Calogero-Moser problem and log-gas in a harmonic field, see \cite{V}. In the case when $k_1,\dots, k_n$ are consequent numbers it can be also interpreted as the pole set of some rational solutions of the fourth Painlev\'e equation and has a regular rectangle-like structure  in the complex plane, as was revealed numerically by Clarkson \cite{C}. A natural question is what kind of pattern do we have in general.

Let us label these potentials by partitions $\lambda=(\lambda_1, \dots, \lambda_n),\, \lambda_1\geq\dots \geq \lambda_n\geq 1,$ such that $\lambda_i=k_{i}-n+i,\, i=1,\dots, n:$
$$k_1=\lambda_1+n-1,\, k_2=\lambda_{2}+n-2,\, \dots,k_{n-1}=\lambda_{n-1}+1,\, k_n=\lambda_n.$$

Our main observation (based on numerical experiments using Mathematica) is that although for a general partition $\lambda$ the picture is quite complicated, for its doubled version $$\lambda^{2\times 2}=((2\lambda_1)^2, \dots, (2\lambda_n)^2)$$ there exists a simple qualitative relation between the shape of the Young diagram and the pattern of zeroes of the corresponding Wronskian. 

In the case of the two-term Wronskian $W(H_n, H_{n+k})$ we have some quantitative results.
Namely, for fixed $k$ and large $n$ we give an explicit formula 
for the curve the scaled zeroes $w=z/\sqrt {2n}=u+iv$ in the region $|u|<1-\delta,\, |v|> \varepsilon \frac{\log n}{n}, \, \varepsilon, \delta >0$ lie on asymptotically :
\begin{equation}
\label{lim}
|v| = \frac{1}{4n\sqrt{1-u^2}} \Big( \ln \big(\frac{8n}{k}\big) + \ln{(1-u^2)} + \frac{1}{2}\ln |1-T_k^2(u)|\Big), 
\end{equation}
where $T_k(x)=\cos k \arccos x$ is the $k$-th Chebyshev polynomial.
The distribution of the real parts of the zeroes on this curve satisfy Wigner's semicircle law.
 The derivation is based on a version of the classical Plancherel-Rotach formula \cite{Sz} found by Deift et al in \cite{DKMVZ}.
 We give also some empirical formulas for the three and four-term Wronskians.

\section{Wronskians of Hermite polynomials and their zeroes}

 Hermite polynomials $H_n(x)$ are the classical orthogonal polynomials with Gaussian weight $w(x)=e^{-x^2}$ (see e.g. \cite{Sz}). They can be given by the formula
 $$ H_n(x)=(-1)^n e^{x^2}\frac{d^n}{dx^n}e^{-x^2}=e^{x^2/2}\bigg (x-\frac{d}{dx} \bigg )^n e^{-x^2/2}$$ 
and satisfy the recurrence relation
$$H_{n+1}(x)=2xH_n(x)-2nH_{n-1}(x).$$ Here are the first few of them:
$$H_0(x)	=	1,\,
H_1(x)	=	2x,\,
H_2(x)	=	4x^2-2,\,
H_3(x)	=	8x^3-12x,$$
$$H_4(x)	=	16x^4-48x^2+12,\,
H_5(x)	=	32x^5-160x^3+120x,...$$
We are using the normalisation where the highest coefficient of $H_n$ is $2^n,$ but this will not be essential in what follows. 
What is important for us is that  $\psi_n=H_n(x) e^{-x^2/2}$ are the eigenfunctions of the harmonic oscillator:
$$ (-\frac{d^2}{dx^2} + x^2)\psi_n= (n+1/2)\psi_n, \,,n=0,1,\dots .$$

Let $\lambda=(\lambda_1, \dots, \lambda_n)$ be a partition and consider the Wronskian
$$W_{\lambda}(z) = W(H_{\lambda_1+n-1}(z), H_{\lambda_{2}+n-2}(z), \dots, H_{\lambda_{n-1}+1}(z), H_{\lambda_n}(z)).$$

The Wronskians $W_{\lambda}$ have the following properties:

\medskip

{\it  1. $W_{\lambda}(z)$ is a polynomial in $z$ of degree $|\lambda|=\lambda_1+\lambda_2+\dots +\lambda_n,$
 
 2. $W_{\lambda}(-z)=(-1)^{|\lambda|} W_{\lambda}(z),$
 
 3. $W_{\lambda^*}(z) = (-i)^{|\lambda|} W_{\lambda}(iz),$ where $\lambda^*$ is the conjugate of $\lambda$.}
 
 \medskip

Recall that the conjugate to a partition $\lambda$ is a new partition, whose Young diagram is the transpose of the diagram of $\lambda.$ To prove the last (duality) property we note that the harmonic oscillator has also the following (growing, hence formal) eigenfunctions
$\psi^*_n=H^*_n(x) e^{x^2/2}, \quad H^*_n(x)=(-i)^n H_n(ix)$
with the negative eigenvalues $\lambda=-n+1/2, \, n=0,1,\dots.$ The claim is that the result of applications of Darboux transformations at the levels $\lambda_i+p-i, \, i=1,\dots, p,$ $p$ is the length of the partition, and at the negative levels $j-q-\lambda^*_j, \, j=1,\dots, q, q$ is the length of the conjugate partition, lead to the potentials different only by a constant shift. This follows, for example, from proposition (1.7) in  Macdonald's book  \cite{Mac}), or from the so-called Maya representation of the partition.

\begin{figure}
\centerline{ \includegraphics[width=5cm]{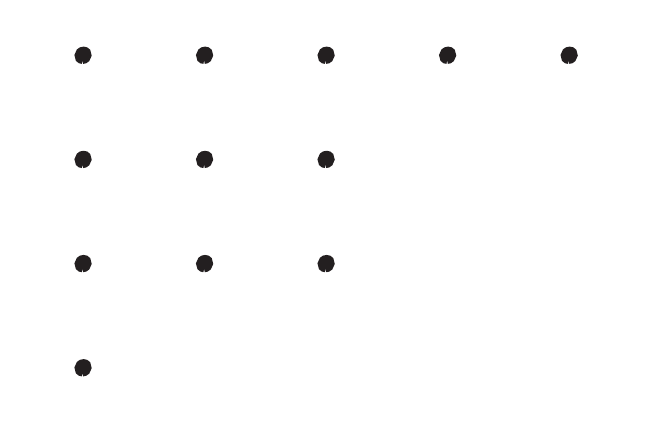} \hspace{20pt} \includegraphics[width=5cm]{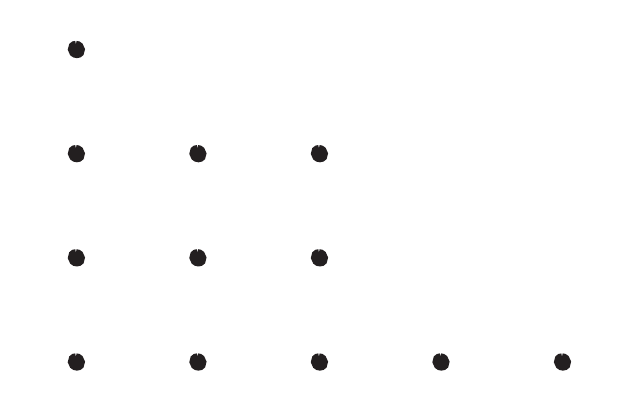} }
\caption{Ferrers diagram for the partition $\lambda=(5,3,3,1).$ Left: standard version. Right: French version} \label{Fer}
\end{figure}

Recall the following  well-known diagrammatic  representations of a partition $\lambda$  (see e.g.\cite{Mac}). 
The first one  is the set of points $(i,j)\in \mathbb Z^2$ such that $1\leq j\leq \lambda_i.$ Following \cite{Sagan} we will call it a {\it Ferrers diagram}. There are two ways to draw this. One convention (motivated by matrix theory) is that the row index $i$ increases downwards while $j$ increases as one goes from left to right. Another way (sometimes called French) is to use a natural Cartesian coordinate representation (see Fig. \ref{Fer}).

\begin{figure}
\centerline{ \includegraphics[width=10cm]{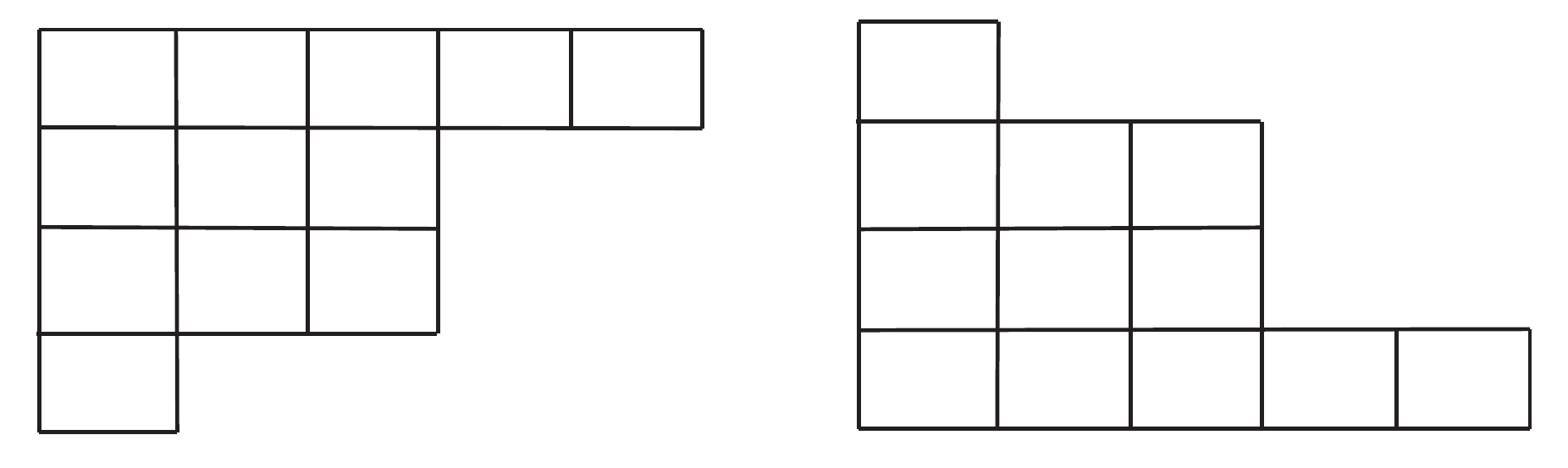} }
\caption{Standard and French versions of the Young diagram for the partition $\lambda=(5,3,3,1)$. } \label{Young}
\end{figure}

The second, most common, way, known as a {\it Young diagram}, is to use boxes rather than bullets (see Fig. \ref{Young}).

\begin{figure}
\centerline{ \includegraphics[width=14cm]{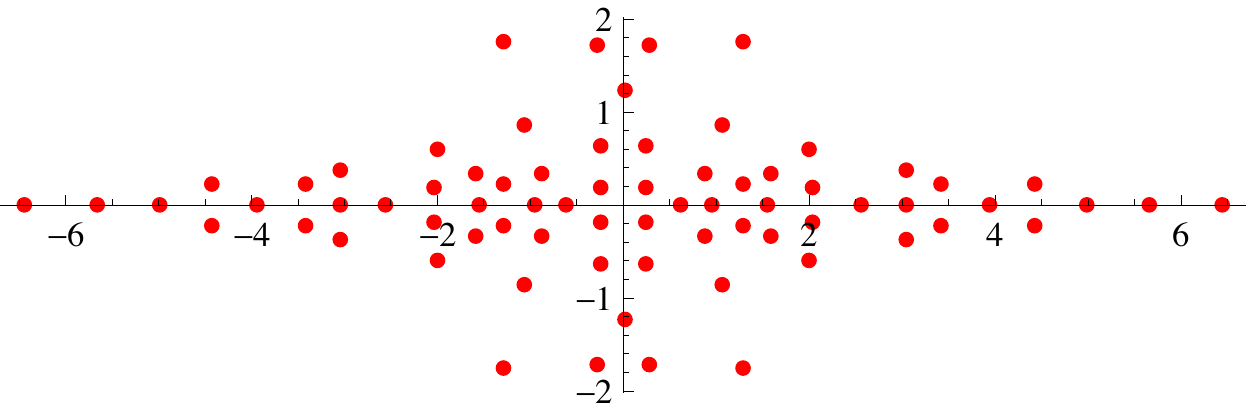} }
\caption{Zeroes of the Wronskian $W_{\lambda}$ with $\lambda=(28,16,10,6,4,4,3,1)$. } \label{comp}
\end{figure}

Since the Wronskians are labelled by the partitions, we can ask a natural question: how is the geometry of the corresponding diagram of $\lambda$ related to the pattern of the zeroes of $W_{\lambda}(z)$ ? Figure \ref{comp}, produced with the help of Mathematica, shows that in general such a relation is not easy to see. Another example is the partition $\lambda=(n,n-1,n-2,\dots, 2,1)$ with a triangular Young diagram, for which the corresponding Wronskian $W_{\lambda}$ (up to a multiple) is simply $z^{n(n+1)/2},$ so we just have one zero at $z=0$ with multiplicity $n(n+1)/2.$

This is why we found it very interesting that for a special class of partitions, which we call doubled,
one can read off the partition from the pattern of zeroes in a straightforward way.

\section{Doubled partitions and their diagrams}

Let $\lambda=(\lambda_1,\lambda_2, \dots, \lambda_n)$ be a partition. Define its {\it doubled version} as
$$\lambda^{2\times 2} = (2\lambda_1, 2\lambda_1, 2\lambda_2, 2\lambda_2, \dots, 2\lambda_n, 2\lambda_n).$$
In other words, we double all parts and take them twice. For example, when $\lambda=(5,3,2)$ the doubled version is $\lambda^{2\times 2} =(10,10,6,6,4,4)=(10^2,6^2,4^2),$ where the power denotes how many times this part is repeated.

\begin{figure}
\centerline{ \includegraphics[width=10cm]{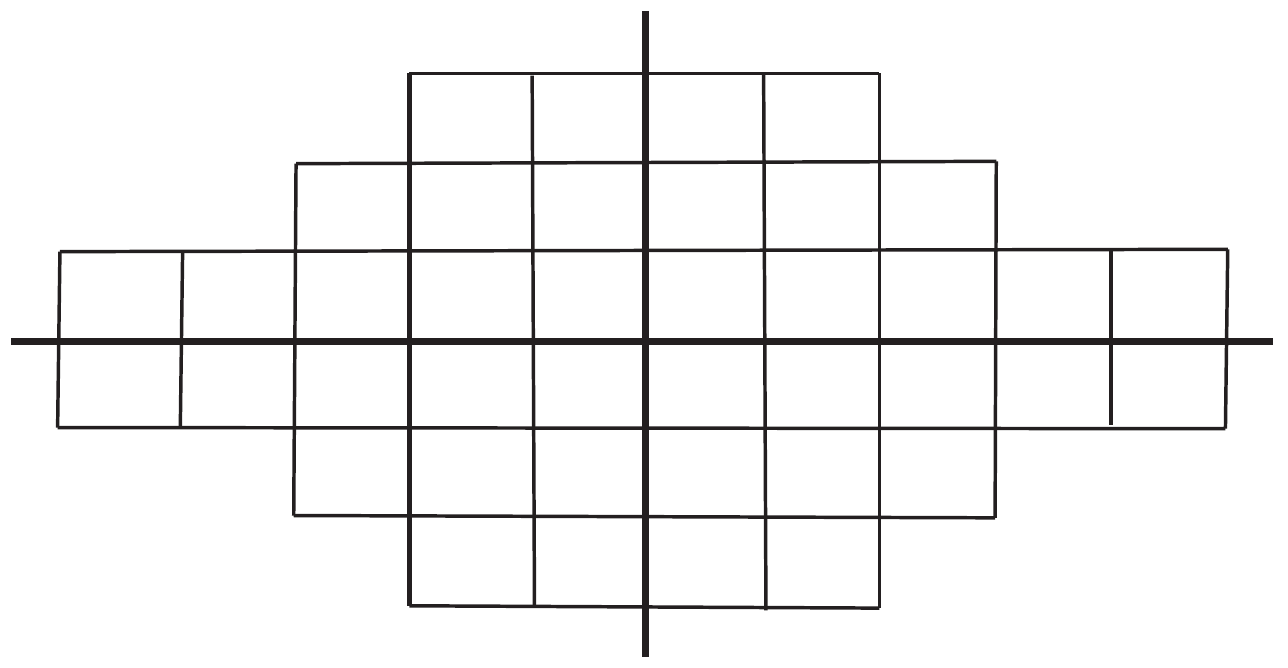} }
\caption{Diagram of the doubled partition $\lambda^{2\times 2}$ for $\lambda=(5,3,2).$} \label{doub}
\end{figure}

Note that the shape of the Young diagram of the doubled version is similar (with scaling factor 2) to the initial one. However, for the doubled partitions there is another natural way to represent them, which combines both the usual and the French ways. Namely one can put the diagram of $\lambda$ in all 4 quadrants as in Fig. \ref{doub}. 
One can naturally define the Ferrers version, which we combine with the Young version by putting bullets at the centre of each box.

\begin{figure}
\centerline{ \includegraphics[width=14cm]{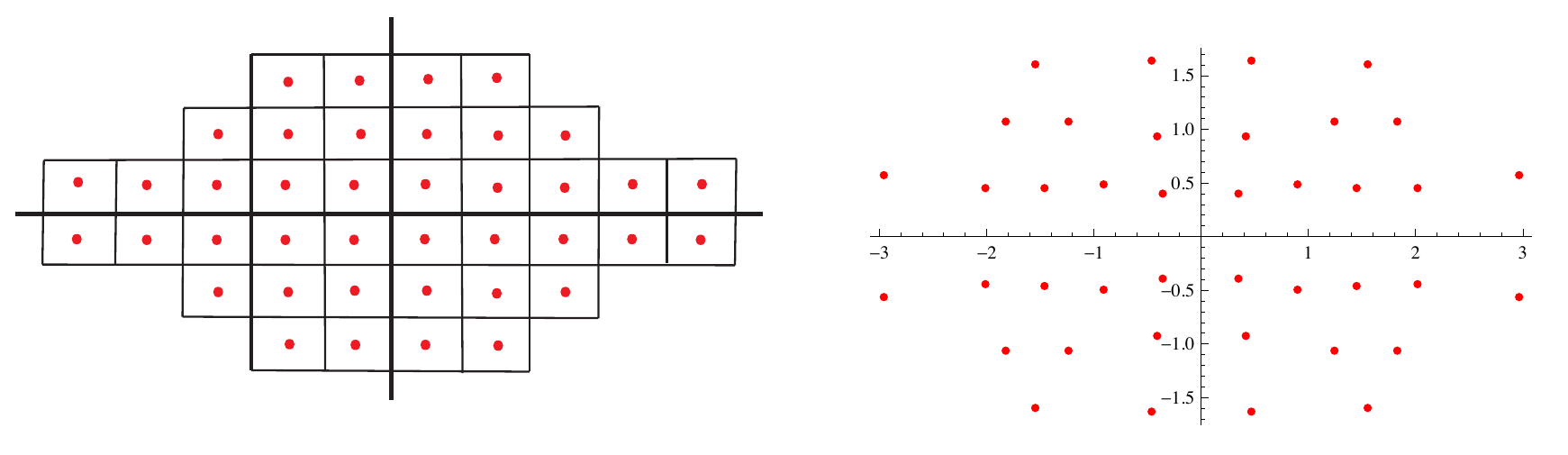} }
\caption{Bulleted diagram of the doubled partition $\lambda^{2\times 2}$ for $\lambda=(5,3,2)$ and the zeroes of the corresponding Wronskian $W_{\lambda^{2\times 2}}$} \label{compa1}
\end{figure}

\begin{figure}
\centerline{ \includegraphics[width=14cm]{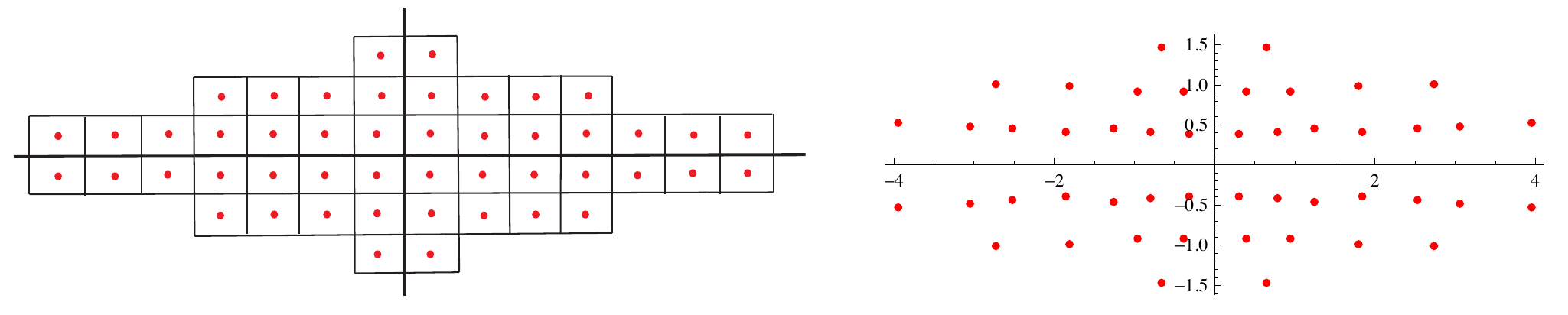} }
\caption{The same comparison for $\lambda=(7,4,1).$} \label{compa2}
\end{figure}

Our main observation is that {\it the diagram of the doubled partition $\lambda^{2\times 2}$ gives a good qualitatative description of the zero set of the corresponding Wronskian} $W_{\lambda^{2\times 2}}$, see Fig. \ref{compa1} and \ref{compa2}.
We believe that this works for any partition $\lambda$ with distinct $\lambda_i.$ 

\begin{figure}[h]
\centerline{ \includegraphics[width=6cm]{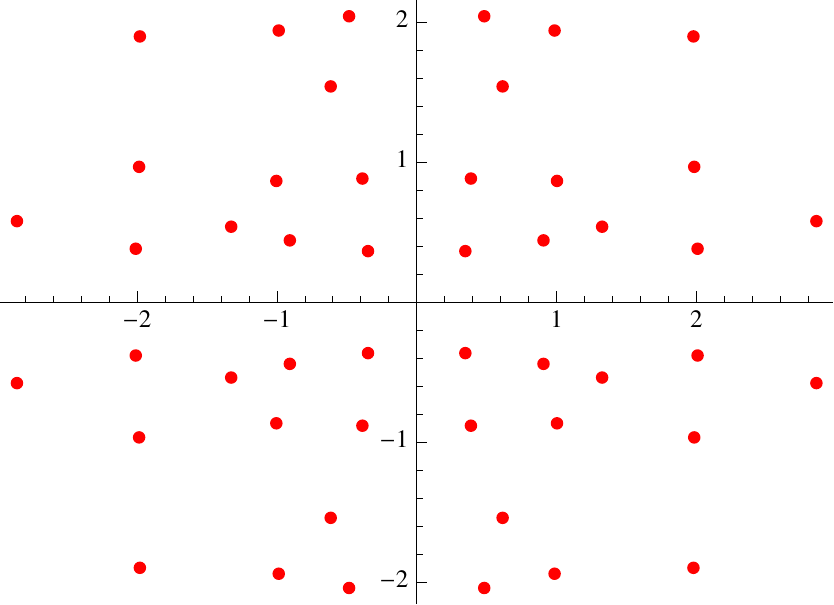} }
\caption{Interference between the rows of zeroes of $W_{\lambda^{2\times 2}}$ for partition $\lambda=(5,3,3,1)$ with two equal parts.} \label{inter}
\end{figure}

When some parts are equal then we may have interference between the rows of corresponding zeroes, see Fig. \ref{inter}. 

\begin{figure}
\centerline{ \includegraphics[width=6cm]{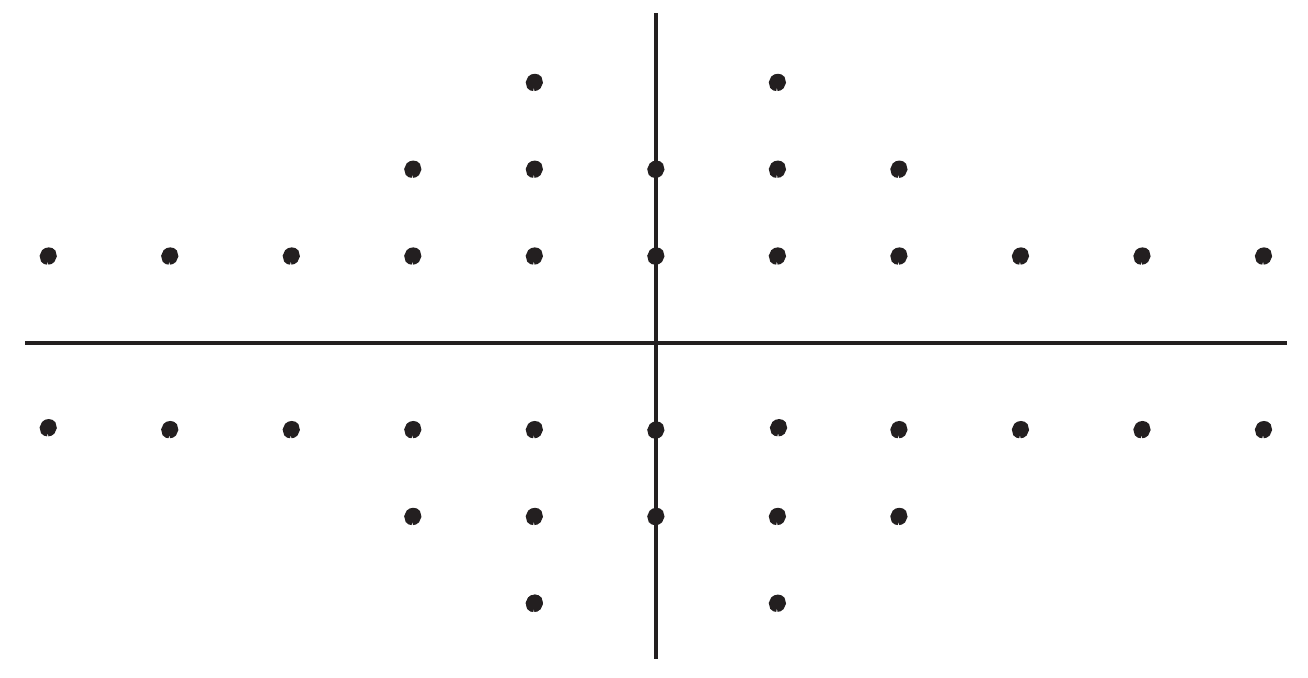} \hspace{20pt} \includegraphics[width=6cm]{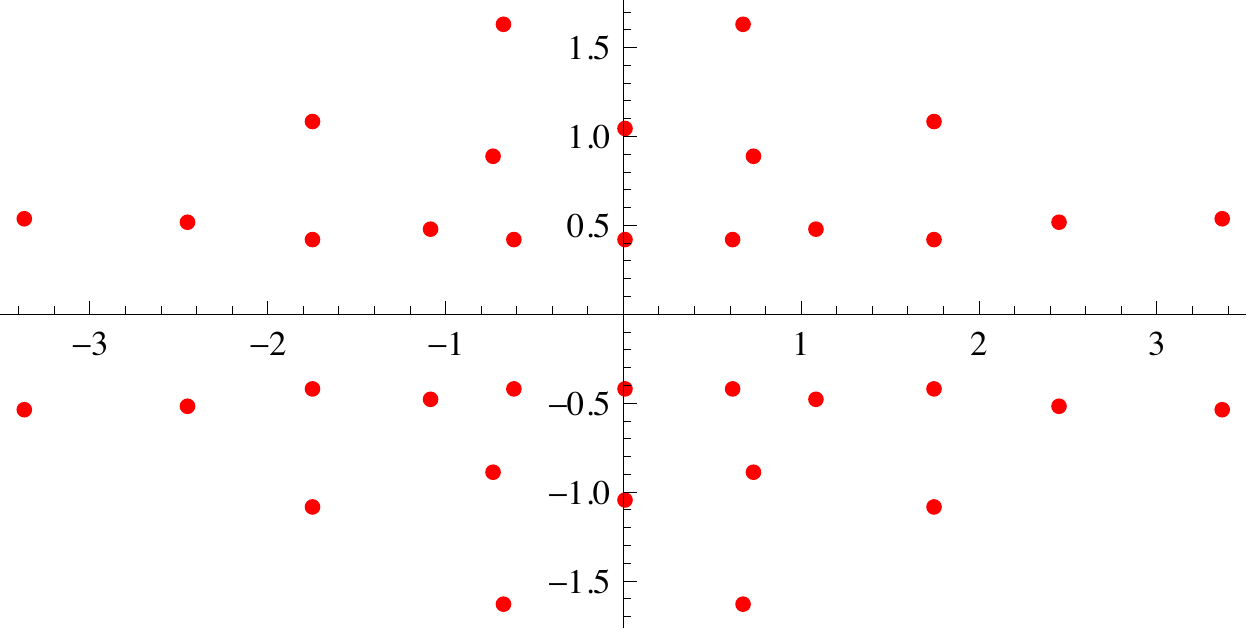}}
\caption{Ferrers diagram (left) of $\lambda^{2\times 2}$ and zeroes of $W_{\lambda^{2\times 2}}$ (right) for half-integer partition $\lambda=(11/2, 5/2, 1)$.} \label{half}
\end{figure}

One can generalise this relation to the case of half-integer partitions $(\lambda_1, \dots, \lambda_n)$ with some of the parts being half-integers.  An example is $\lambda=(1,5/2,11/2)$ for which the doubled partition is $\lambda^{2\times 2}=(2^2,5^2,11^2).$ The corresponding Ferrers diagram has some bullets on the vertical axis and gives a  good qualitative picture of the zero set of the corresponding Wronskian (see Fig. \ref{half}). 

The following asymptotic analysis shows the limits of this comparison already in the simplest case of one-row Young diagram $\lambda$.

\section{Asymptotic behaviour of zeroes of two-term Wronskians}

Consider now the two-term Wronskian $W(H_n, H_{n+k})$, corresponding to the partition $\lambda=(n+k-1,n), k \geq 1.$
Let us fix $k$ and let $n \rightarrow \infty.$ To study this behaviour of zeroes in this limit we can use the following version of Plancherel-Rotach formula due to Deift et al \cite{DKMVZ}. \footnote{We are very grateful to Ken McLaughlin for attracting our attention to this important paper during "Dubrovin-60" conference at Sardinia in June 2010.} 
 
In the scaled variable $w=z/\sqrt{2n}$ there are several regions with different asymptotic behaviour of the Hermite polynomials (see Fig. 9). The most relevant for us is the region $B_{\delta},$ where we have the following asymptotics 

\begin{figure}[h]
\centerline{ \includegraphics[width=8cm]{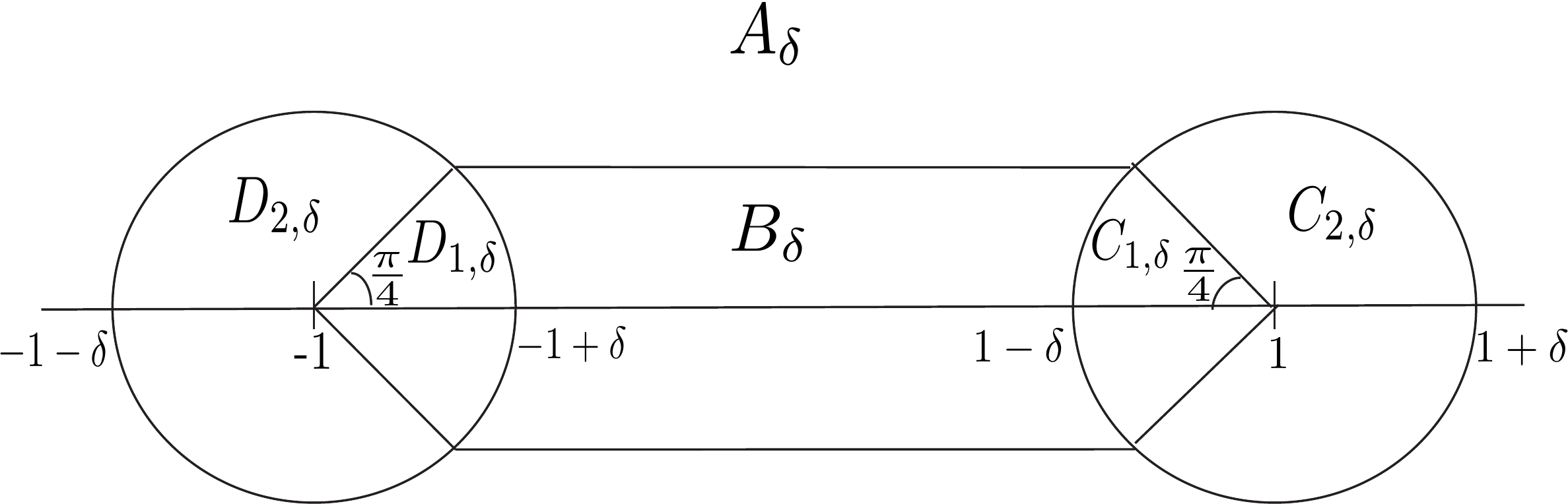}}
\caption{Asymptotic regions in scaled variable $w$} 
\end{figure}

$$
H_n(z)e^{-\frac{z^2}{2}}  = C_n(1-w^2)^{-\frac{1}{4}}\Big( \cos(2n\Theta(w) + \chi(w))(1+O(\frac{1}{n})) + \sin(2n\Theta(w)-\chi(w))O(\frac{1}{n})\Big) 
$$
with $C_n= \sqrt{\frac{2}{\pi}}(2n)^{-\frac{1}{4}}$, $\Theta(w) = \frac{1}{2}w\sqrt{1-w^2} + \frac{1}{2} \arcsin{w} - \frac{\pi}{4}$ and $\chi(w) = \frac{1}{2}\arcsin w$ (see \cite{DKMVZ}).

Using this we can show that $$W(H_n, H_{n+k})=-\frac{2}{\pi}\Big[ \big(\sin\Delta_k + \frac{1}{4n(1-w^2)} (\sin \Delta_k+k \sin(2\Phi+\Delta_k)\big)\Big] (1+O(\frac{1}{n})),$$ 
where $\Phi = 2n \Theta + \chi,\, \Delta_k (w)  =  2k\Theta (w) - kw \Theta'(w) = k \arccos w.$
In the upper half-plane we have two competing terms: $\sin\Delta_k$ term and the negative exponential component $e^{-i(2\Phi + \Delta_k)}$ of $ \sin(2\Phi+\Delta_k)$. 
For $w = u+iv$ with small $v <<1$ we can approximate $\Phi(w)$ as 
$$\Phi(u+iv) \approx 2n\Theta(u) + 2niv\Theta'(u) = 2n\Theta(u) + 2niv\sqrt{1-u^2}$$ since $\Theta'(w) = \sqrt{1-w^2}.$
Equating the moduli of the two competing terms, we have
$$
\Big|\sin\Delta_k(u) \Big|= \frac{k}{8n(1-u^2)}e^{4nv\sqrt{1-u^2}},
$$
or,
\begin{equation}
v= \frac{1}{4n\sqrt{1-u^2}} \Big( \ln \big(\frac{8n}{k}\big) + \ln{(1-u^2)} + \ln |\sin (k\arccos u)| \Big). \label{zeroline} \\
\end{equation}
This leads to the formula (\ref{lim}) for the curve on which the scaled zeroes lie asymptotically as $n\rightarrow \infty$ in the region $|u|<1-\delta,\, |v|> \varepsilon \frac{\log n}{n}, \, \varepsilon, \delta >0.$ Comparing the arguments of the leading terms  and using the relation
$d \Theta =  \sqrt{1 - w^2} dw$ we see that the real parts of the corresponding zeroes are distributed according to the famous {\it Wigner's semicircle law} from the random matrix theory \cite{Wig}: the number $N_{\alpha, \beta}(n, k)$ of scaled zeroes $w=u+iv$ of $W(H_n, H_{n+k})$ in the upper-half plane with the real parts in the interval $(\alpha, \beta)$ satisfies
\begin{eqnarray}
\label{Wigner}
\lim_{n \to \infty} \frac{N_{\alpha, \beta}(n, k)}{n}  =  \frac{2}{\pi} \int_{\alpha}^{\beta} \sqrt{1 - u^2} du.
\end{eqnarray}

When $v=0$ we have $k-1$ real (scaled) zeroes $u_m=x_m/\sqrt{2n}$ asymptotically given by $1-T_k^2(u)=0$: as $n \rightarrow \infty$ 
\begin{equation}
\label{asym2}
u_m=\frac{x_m}{\sqrt {2n}} \rightarrow \cos \frac{\pi m}{k}, \quad m=1,\dots, k-1.
\end{equation}

The unscaled zeroes $z=x+iy$ in the region $$\Omega_{\varepsilon, \delta}: |x|<(1-\delta) \sqrt{2n},\, |y|> \varepsilon \frac{\log n}{\sqrt n}$$  lie asymptotically on the curve
\begin{equation}
\label{asym}
|y| = \frac{1}{2\sqrt{2n-x^2}} \Big( \ln \big(\frac{8n}{k}\big) + \ln{(1-\frac{x^2}{2n})} + \frac{1}{2}\ln |1-T_k^2(\frac{x}{\sqrt{2n}})|\Big), 
\end{equation}
where $T_k(x)$ is the $k$-th Chebyshev polynomial.
Figure 10 shows a good agreement of this formula (curve) and numerical Mathematica calculation of zeroes (dots) in the case when $n=100, k=5.$
The 4 real zeroes in this case approximately are
$$x\approx \pm \sqrt {200} \, \cos \frac{\pi m}{5}=5 \sqrt {2} \, \frac{\pm 1 \pm \sqrt 5}{2}, \quad m=1, 2$$
in agreement with the picture.

\begin{figure}[htp]
\centerline{ \includegraphics[width=10cm]{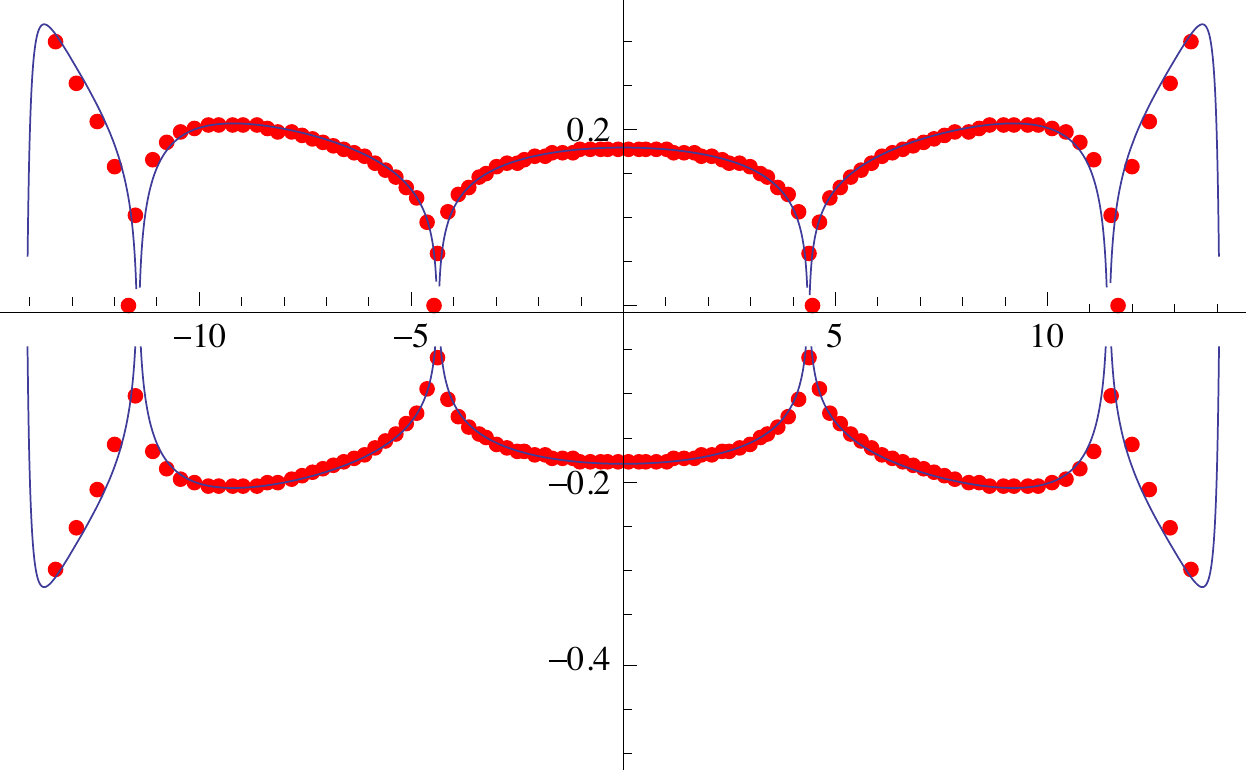}}
\caption{Comparison in the case n=100, k=5} 
\end{figure}

The case $n=100, k=1$ corresponds to the doubled partition $(100, 100)=50^{2\times 2}$.
Figure 11 shows that shapes of the Young diagram and the corresponding zero set coincide only qualitatively. Indeed, the corresponding asymptotic curve in this case is not just two straight lines but given by (\ref{asym}) with $k=1$:
$$
|y| = \frac{1}{2\sqrt{2n-x^2}} \Big( \ln \big(8n\big) + \frac{3}{2}\ln{(1-x^2/2n)}\Big). 
$$

\begin{figure}[htp]
\centerline{ \includegraphics[width=8cm]{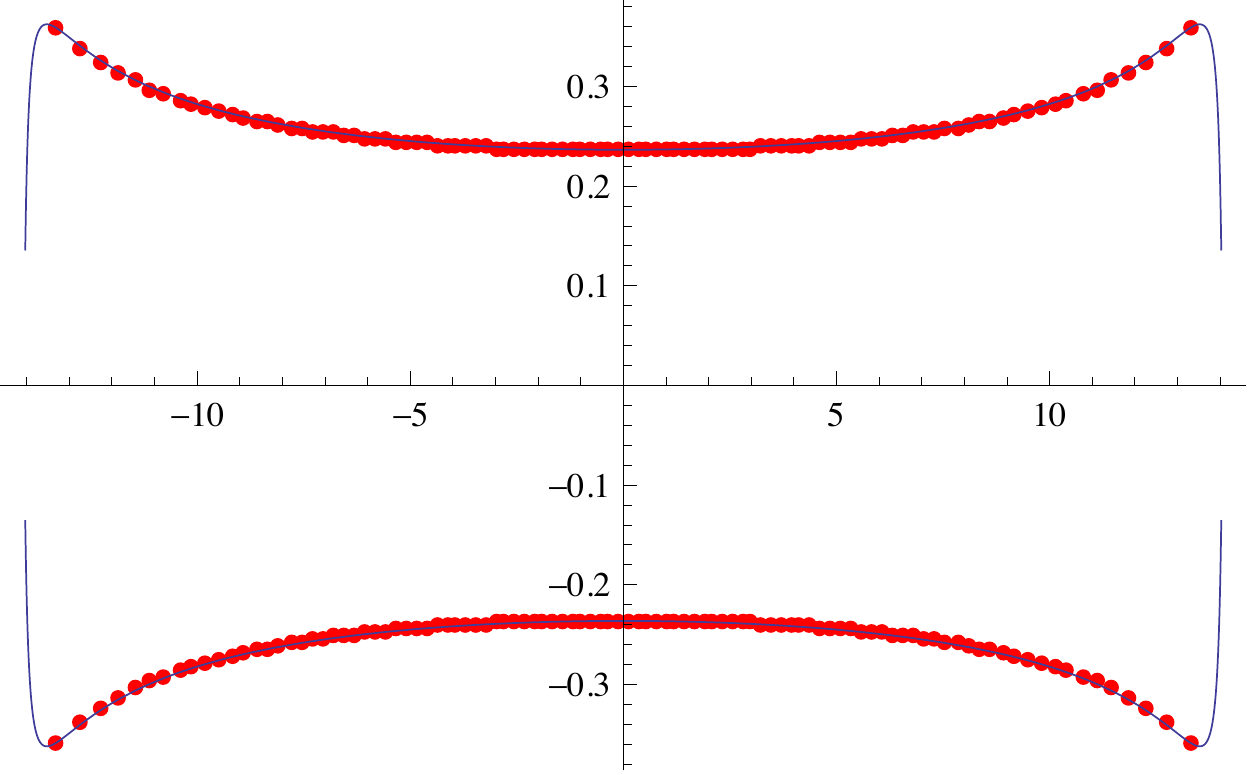}}
\caption{Zeroes of $W(H_{100}, H_{101})$} 
\end{figure}

\section{Empirical asymptotic formulas for 3-term and 4-term Wronskians}

\subsection{Three-term Wronskians. } One can prove the following identity for 3-term Wronskians of the eigenfunctions of harmonic oscillator
$$
W3= W(\psi_{n-k}, \psi_{n}, \psi_{n+l}) = k \psi_{n-k} W(\psi_{n},\psi_{n+l}) - l \psi_{n+l}W(\psi_{n-k},\psi_{n}). 
$$
We will restrict ourselves with the case $k=l.$ In that case we have
\begin{equation}
\label{W3*}
W3= k (\psi_{n-k}W(\psi_{n},\psi_{n+k}) - \psi_{n+k} W(\psi_{n-k},\psi_{n})) 
\end{equation}

Using this, previous results about 2-term Wronskians and experimenting with Mathematica we can suggest the following empirical formula for the limiting shape of the non-real zeroes for large $n$ and $k<<n$:
\begin{equation}
\label{W3as}
|y|= \frac{1}{\sqrt{2n-x^2}} \Big( \ln \big(\frac{6n}{k}\big) + \ln{(1-x^2/2n)} + \frac{1}{2}\ln |1- T_k^2( x/\sqrt{2n})|\Big) \\
\end{equation}
Mathematica plots of zeroes against the corresponding curve below show a good agreement with this formula.

\begin{figure}[h]
\centerline{ \includegraphics[width=6cm]{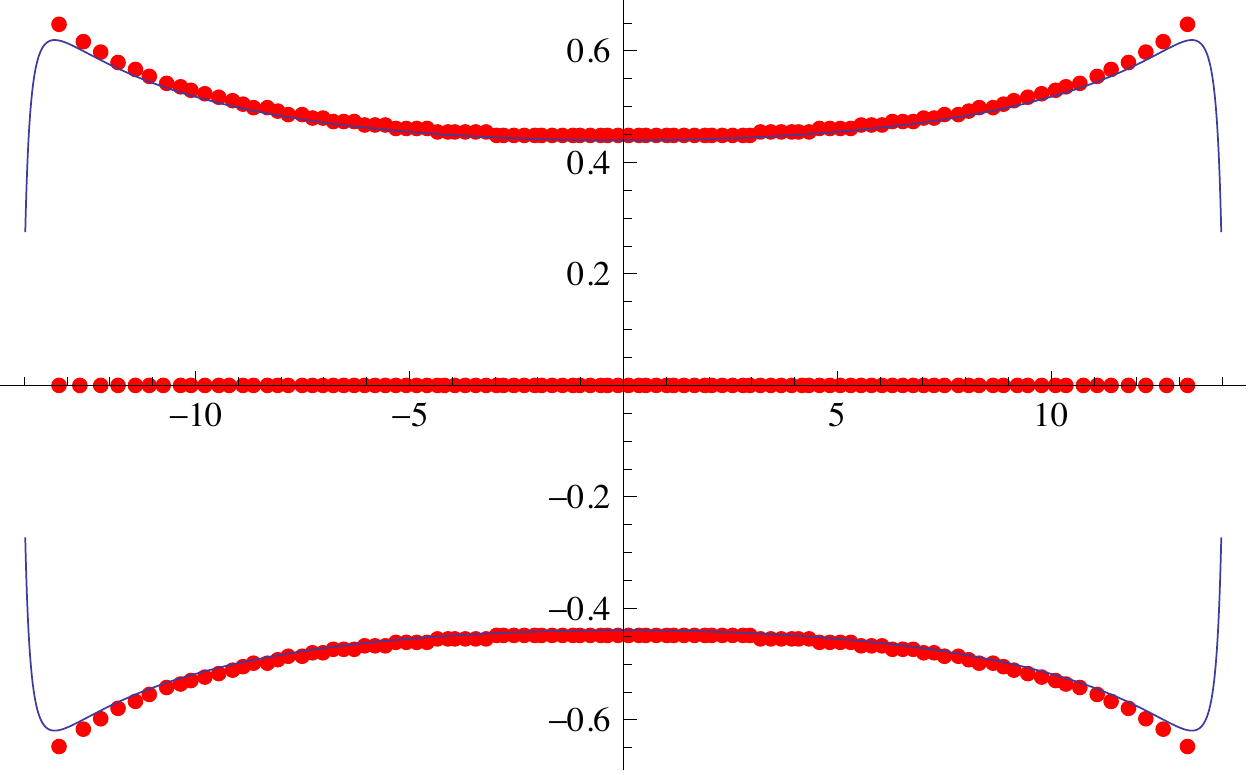}  \hspace{2pt}  \includegraphics[width=6cm]{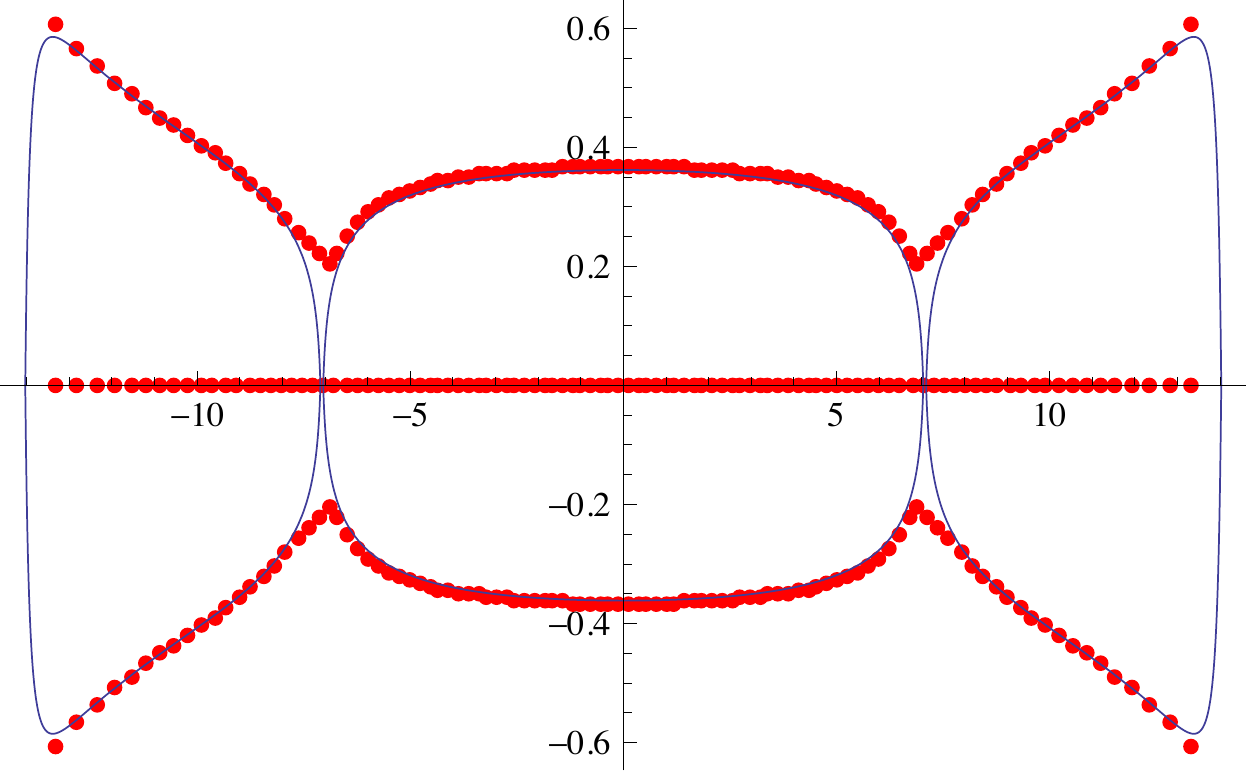}}
\caption{Comparison of the zeroes and the curve for Left:$W(H_{100},H_{101},H_{102})$ Right:$W(H_{100},H_{103},H_{106})$.} 
\end{figure}

\begin{figure}[h]
\centerline{ \includegraphics[width=9cm]{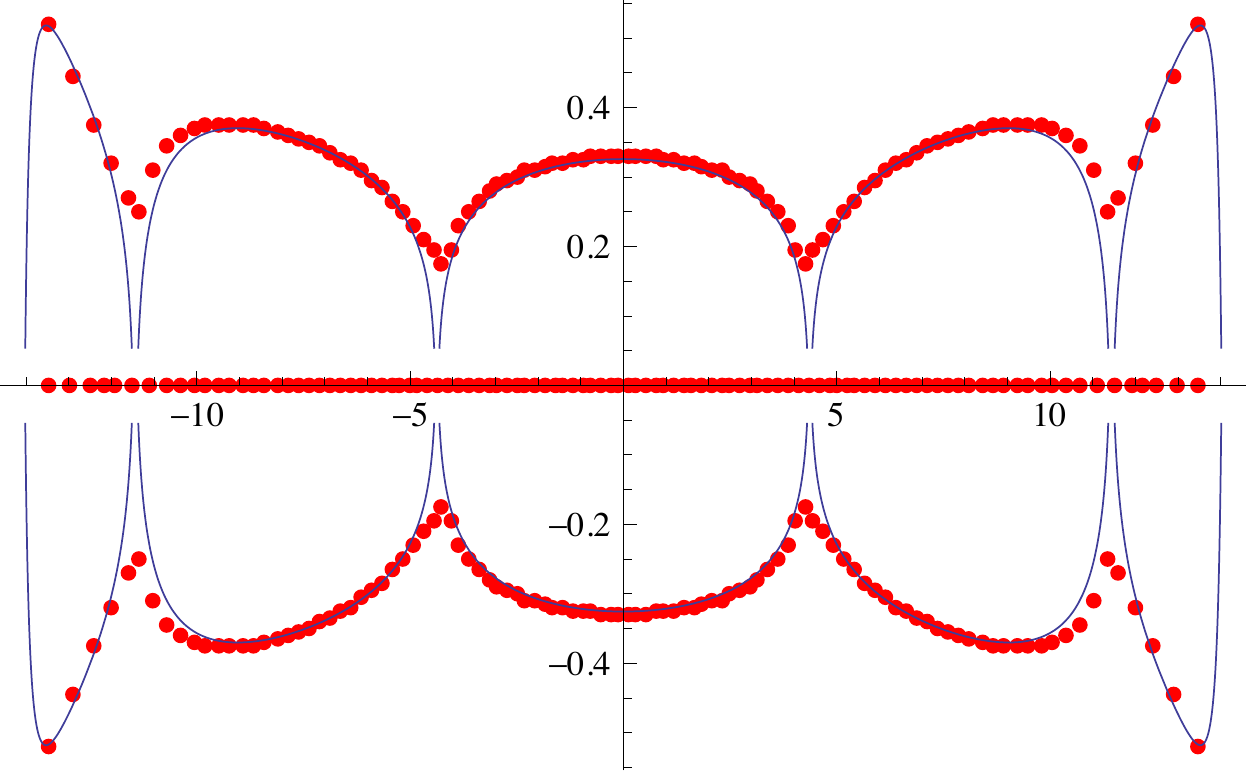} }
\caption{Comparison of the zeroes and the curve for $W(H_{100},H_{105},H_{110})$ .} 
\end{figure}

\subsection{4-term Wronskians. } 
For Wronskians $W4=W(\psi_n, \psi_{n+k}, \psi_{n+k+l},\psi_{n+k+l+m})$ one can show that
\begin{eqnarray}
W4	& = & l(k+l+m) W(\psi_n,\psi_{n+k}) W(\psi_{n+k+l},\psi_{n+k+l+m}) \\
& & \quad - km W(\psi_n,\psi_{n+k+l+m})W(\psi_{n+k},\psi_{n+k+l}) \nonumber
\end{eqnarray}

Assuming first that $k=l=m$ we have
\begin{eqnarray}
W4 & = & 3k^2 W(\psi_n,\psi_{n+k}) W(\psi_{n+2k},\psi_{n+3k}) \\
& & \quad - k^2 W(\psi_n,\psi_{n+3k})W(\psi_{n+k},\psi_{n+2k})\nonumber
\end{eqnarray}

Mathematica plots suggest that the zeroes for large $n$ and small $k<<n$ asymptotically lie on two curves, for which we have the following empirical formulas:
\begin{equation}
\label{W4}
	|y|= \frac{1}{2\sqrt{2n-x^2}} \Big( \ln \big(\frac{4n}{k}\big) + \ln{(1-x^2/2n)} +\frac{1}{2} \ln |1- T_k^2( x/\sqrt{2n})|\Big) 
\end{equation}
for the middle curve and
\begin{equation}
\label{W4*}
	|y|= \frac{3}{2\sqrt{2n-x^2}} \Big( \ln \big(\frac{4n}{k}\big) + \ln{(1-x^2/2n)} + \frac{1}{2}\ln |1- T_k^2( x/\sqrt{2n})|\Big) 
\end{equation}
for the outside curve.

Below we compare these curves and Mathematica plots of zeroes for $n=100$ and values of $k$ ranging from 1 to 4. The first picture corresponds to the double partition $\lambda^{2\times 2}$ with $\lambda=(50, 50).$ In all cases we expect the real parts of properly scaled zeroes to satisfy Wigner's semicircle law. We note a peculiar behaviour of zeroes near the points where $1- T_k^2( x/\sqrt{2n})=0$, which needs a special investigation.
 
\begin{figure}[h]
\centerline{ \includegraphics[width=7cm]{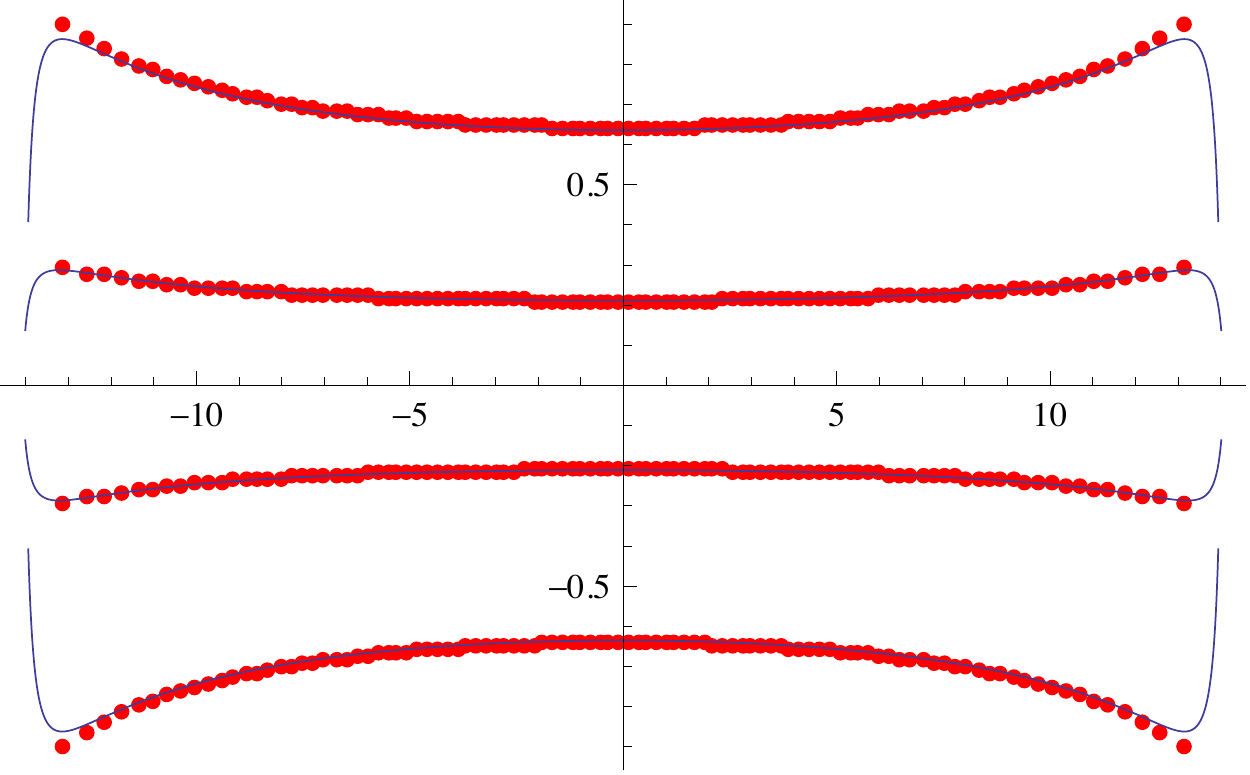}  \hspace{10pt}  \includegraphics[width=7cm]{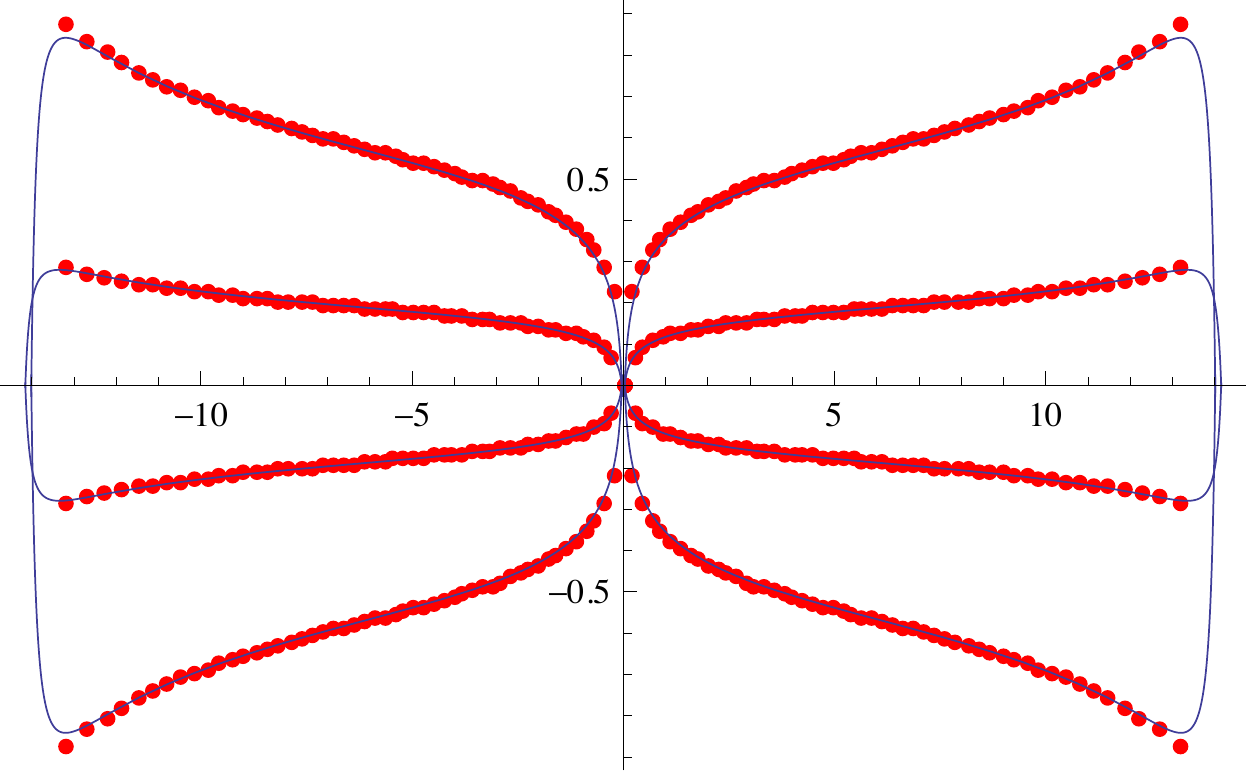}}
\caption{Comparison of the zeroes and the curve for Left:$W(H_{100},H_{101},H_{102},H_{103})$ Right:$W(H_{100},H_{102},H_{104},H_{106})$.} 
\end{figure}
 
\begin{figure}[h]
\centerline{ \includegraphics[width=7cm]{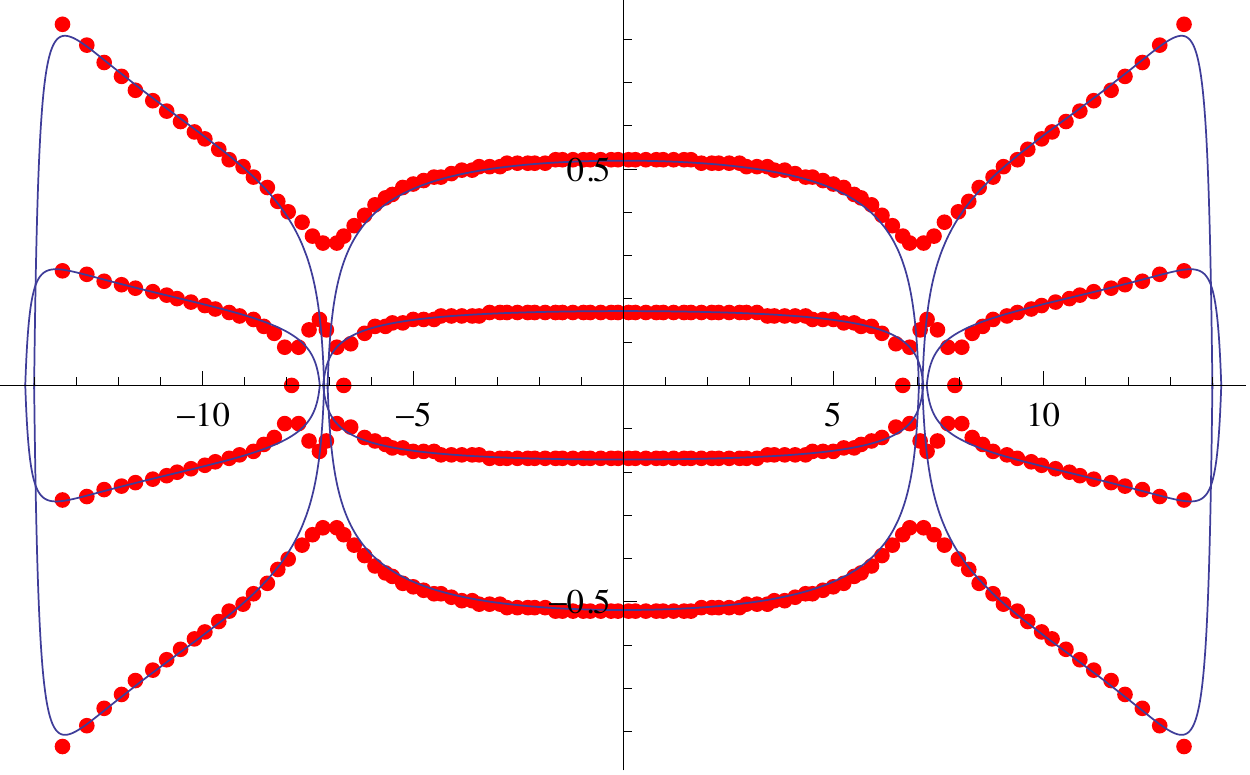}  \hspace{10pt}  \includegraphics[width=7cm]{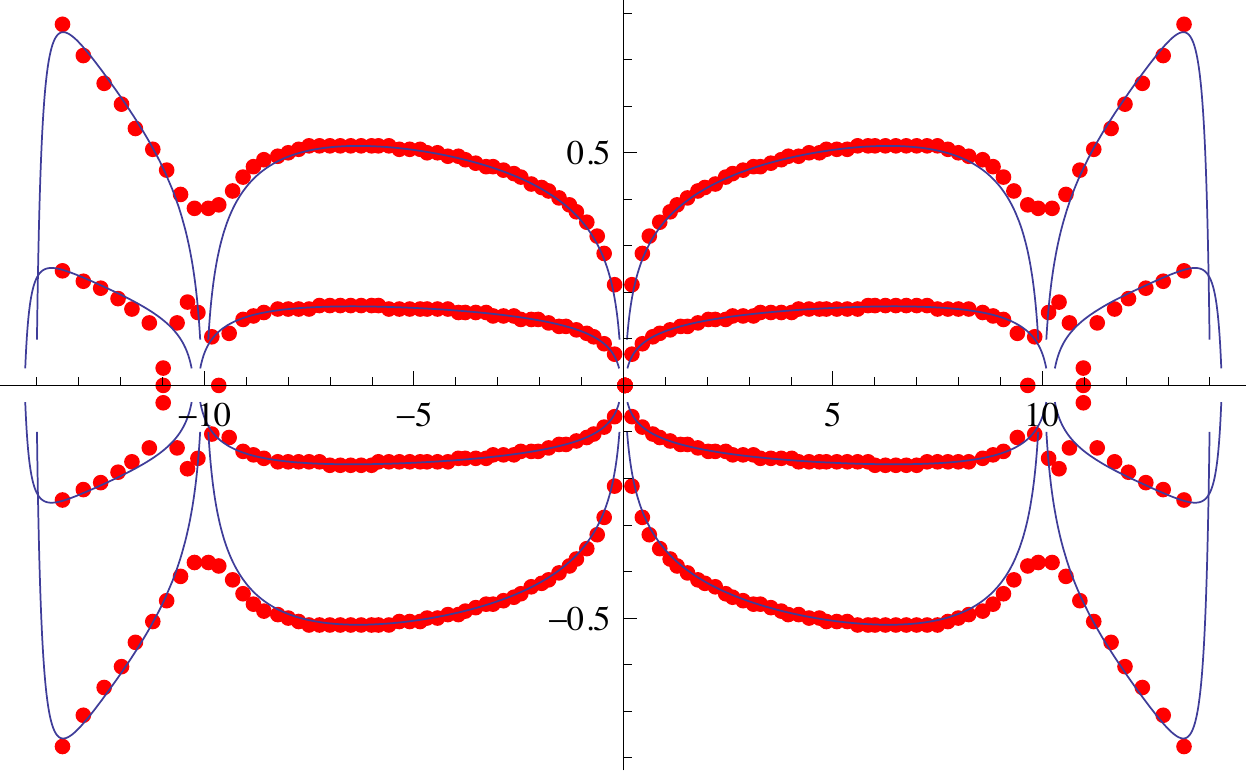}}
\caption{Comparison of the zeroes and the curve for Left:$W(H_{100},H_{103},H_{106},H_{109})$ Right:$W(H_{100},H_{104},H_{108},H_{112})$.} 
\end{figure}

Consider now the case $k=m=1$ and $W4= W(H_n, H_{n+1}, H_{n+l+1}, H_{n+l+2})$, corresponding to the doubled partitions $\lambda^{2\times 2}$ with $\lambda=(n/2, (l-1)/2).$ The empirical formulas for the asymptotic zero curves for large $n$ are

\begin{equation}
\label{W4doub}
	|y|= \frac{1}{2\sqrt{2n-x^2}} \Big( \ln 4n +  \frac{3}{2}\ln{(1-x^2/2n)} -\frac{1}{2} \ln |1- T_l^2( x/\sqrt{2n})|\Big) 
\end{equation}
for the middle curve and
\begin{equation}
\label{W4doub*}
	|y|= \frac{1}{\sqrt{2n-x^2}} \Big( \ln \frac{8n^2}{5l} +  \frac{3}{2}\ln{(1-x^2/2n)} + \frac{1}{l}\ln |1- T_l^2( x/\sqrt{2n})|\Big) 
\end{equation}
for the outside curve. They seem to work fairly well when $l <n/4$ but not too small. The cases with $n=50$, $l=11$ and $n=60, l=10$ are shown below.
We should say that at the moment this part is just experimental mathematics and requires further investigation.

\begin{figure}[h]
\centerline{ \includegraphics[width=10cm]{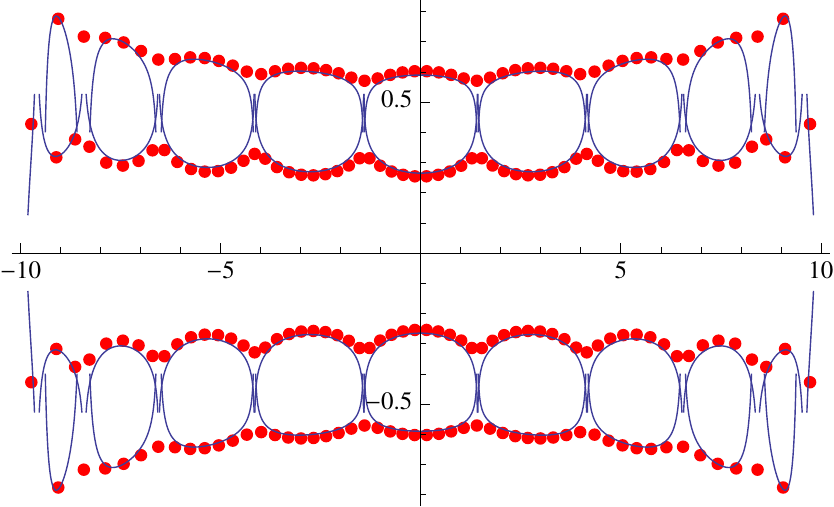} }
\caption{Comparison of the zeroes and the curves for $W(H_{50},H_{51},H_{62},H_{63})$.} \label{Fig:doubled50}
\end{figure}

\begin{figure}[h]
\centerline{ \includegraphics[width=10cm]{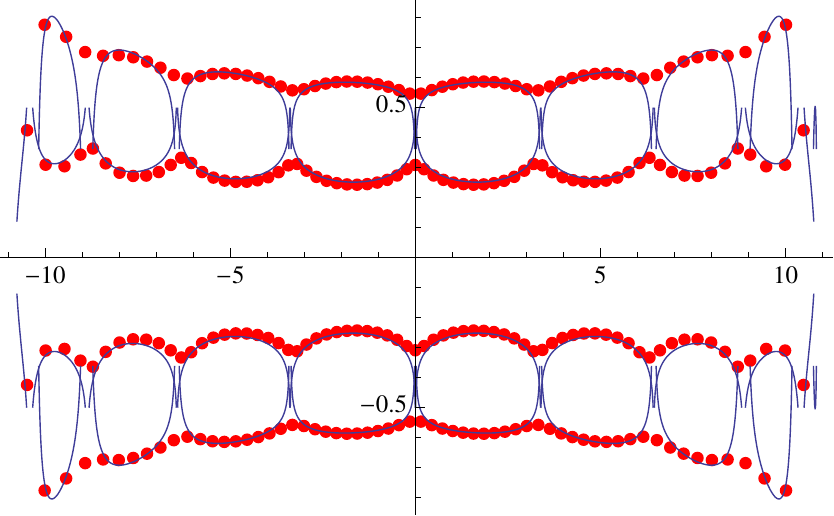} }
\caption{Comparison of the zeroes and the curves for $W(H_{60},H_{61},H_{71},H_{72})$.} \label{Fig:doubled60}
\end{figure}

\section{Some conjectures}

The following property of the Wronskians of Hermite polynomials was conjectured by the third author in 1990s in relation with the corresponding locus problem solved by Oblomkov \cite{O}.
If this property holds it would give a way to a more effective proof of his result, which is still very desirable.

\begin{conj}
For every partition $\lambda$, all the zeroes of $W_{\lambda}(z)$ are simple except possibly for $z=0.$
\end{conj}

Note that the multiplicity $m$ of $z=0$ for $W_{\lambda}$ can be easily computed and has the form
$$m=\frac{d(d+1)}{2},$$
where $d=p-q$ is the difference between the numbers $p$ and $q$ of odd and even elements respectively among the sequence $\lambda_1+n-1, \lambda_2+n-2,\dots, \lambda_{n-1}+1, \lambda_n.$
In particular, for the triangular Young diagram $\lambda=(n, n-1, \dots, 2, 1)$ we have $d=n$ and $m=n(n+1)/2 = \deg W_{\lambda}$, so the corresponding Wronskian $W_{\lambda}=C_n z^{n(n+1)/2}$ and all the zeroes collide at zero.

An interesting question is if the number of real zeroes of $W_{\lambda}(z)$ can be effectively described in terms of the corresponding Young diagram. For doubled partitions we have the following conjecture.

\begin{conj}
For doubled partitions $\nu=(\mu_1^2, \dots, \mu_n^2)$ with distinct parts, the Wronskian $W_{\nu}(z)$ has no real roots and has as many pure imaginary roots as there are odd numbers among $\mu_1,\dots,\mu_n.$
\end{conj}

In the special case when $n=1$ and $\nu=(m,m)$ we can prove this using the integral representation of the corresponding Wronskian known from the random matrix theory (see Br\'ezin-Hikami \cite{BH}):
$$W_{\nu}(z)=c_{m} \int_{-\infty}^{\infty}\dots \int_{-\infty}^{\infty} \prod_{i<j}^m (x_i-x_j)^2 \prod_{k=1}^m (z-x_k)^2 e^{-x_k^2} dx_1\dots dx_m.$$

Finally, it would be very interesting to understand how special the Hermite polynomials are and how much of this can be generalised to other orthogonal polynomials and to the sextic growth case \cite{GV}.

\section{Acknowledgments}

One of us (APV) is grateful to the Institute for Mathematical Research at ETH Zurich for the hospitality in April 2010 and to Robert Milson for stimulating discussions.

\end{document}